\documentclass[prc,aps,fleqn,twoside,11pt]{revtex4} 
\usepackage[dvipdfm]{graphicx}% Include figure files 
\usepackage{dcolumn}% Align table columns on decimal point 
\usepackage{bm}% bold math \usepackage{amsmath} 
\newcommand{\be}{\begin{equation}}
\newcommand{\ee}{\end{equation}}
\newcommand{\bea}{\begin{eqnarray}}
\newcommand{\eea}{\end{eqnarray}}

\newcommand{\Tr}{\mbox{Tr}}
\newcommand{\VEV}[1]{\left\langle #1\right\rangle}

\newcommand{\pbp}{\bar\psi\psi}

\begin{document}
\bibliographystyle{h-physrev}
%\draft

\author{Carleton DeTar}
\affiliation{Department of Physics and Astronomy, University of Utah, Salt Lake City, Utah 84112, USA
\footnote{Temporary address: Center for Computational Sciences, 
Tsukuba University, Tsukuba, Japan}}

\title{QCD Thermodynamics on the Lattice: Recent Results}

\begin{abstract}
I give a brief introduction to the goals, challenges, and technical
difficulties of lattice QCD thermodynamics and present some recent
results from the HotQCD collaboration for the crossover temperature,
equation of state, and other observables.
\end{abstract}

\maketitle

\section{Introduction}

Numerical simulations and models have established that the
high-temperature behavior of QCD at low baryon number density is
governed by two interrelated phenomena, namely the transition from a
low temperature, confined regime to a high temperature deconfined
regime and the transition from a low temperature regime with
spontaneously broken chiral symmetry to a high temperature regime in
which the chiral symmetry is restored.  The deconfinement phenomenon
is especially apparent at very large quark masses where the first
order phase transition of pure SU(3) Yang-Mills theory becomes
manifest.  The chiral restoration phenomenon, on the other hand, is
most relevant in the limit of vanishing quark masses.  Between these
extremes only a nonperturbative calculation can say what happens.  The
present consensus in lattice QCD is that there is no phase transition
--- only a crossover --- at physical quark masses and zero baryon
number density \cite{Hwa:2010,Yagi:2008}.

\begin{figure}
  \begin{tabular}{lr}
    \begin{minipage}{0.415\textwidth}
  \vspace*{-2mm}
  \includegraphics[width=\textwidth]{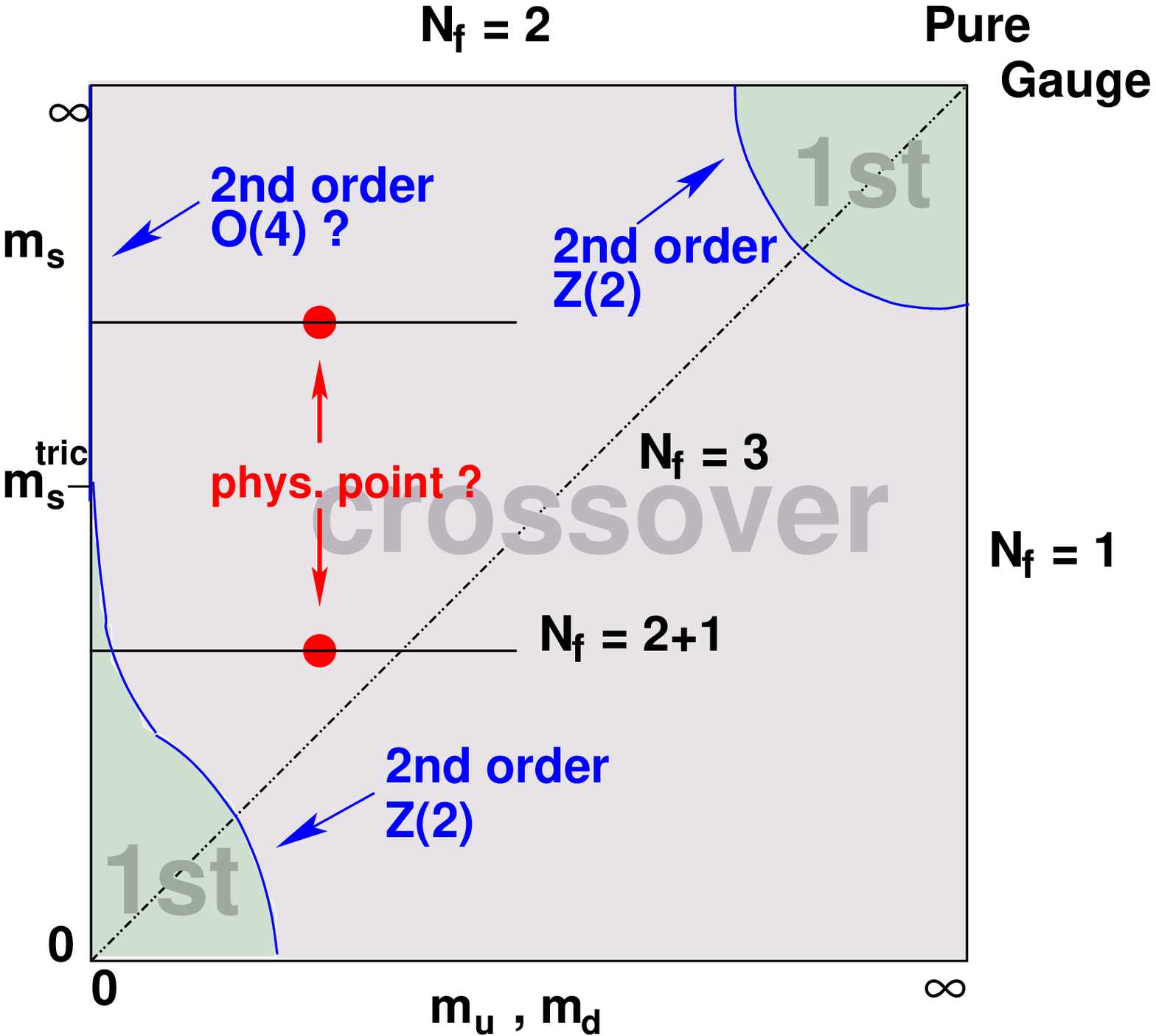}\hfill
    \end{minipage}
&
    \begin{minipage}{0.385\textwidth}
\vspace*{3mm}
\includegraphics[width=\textwidth]{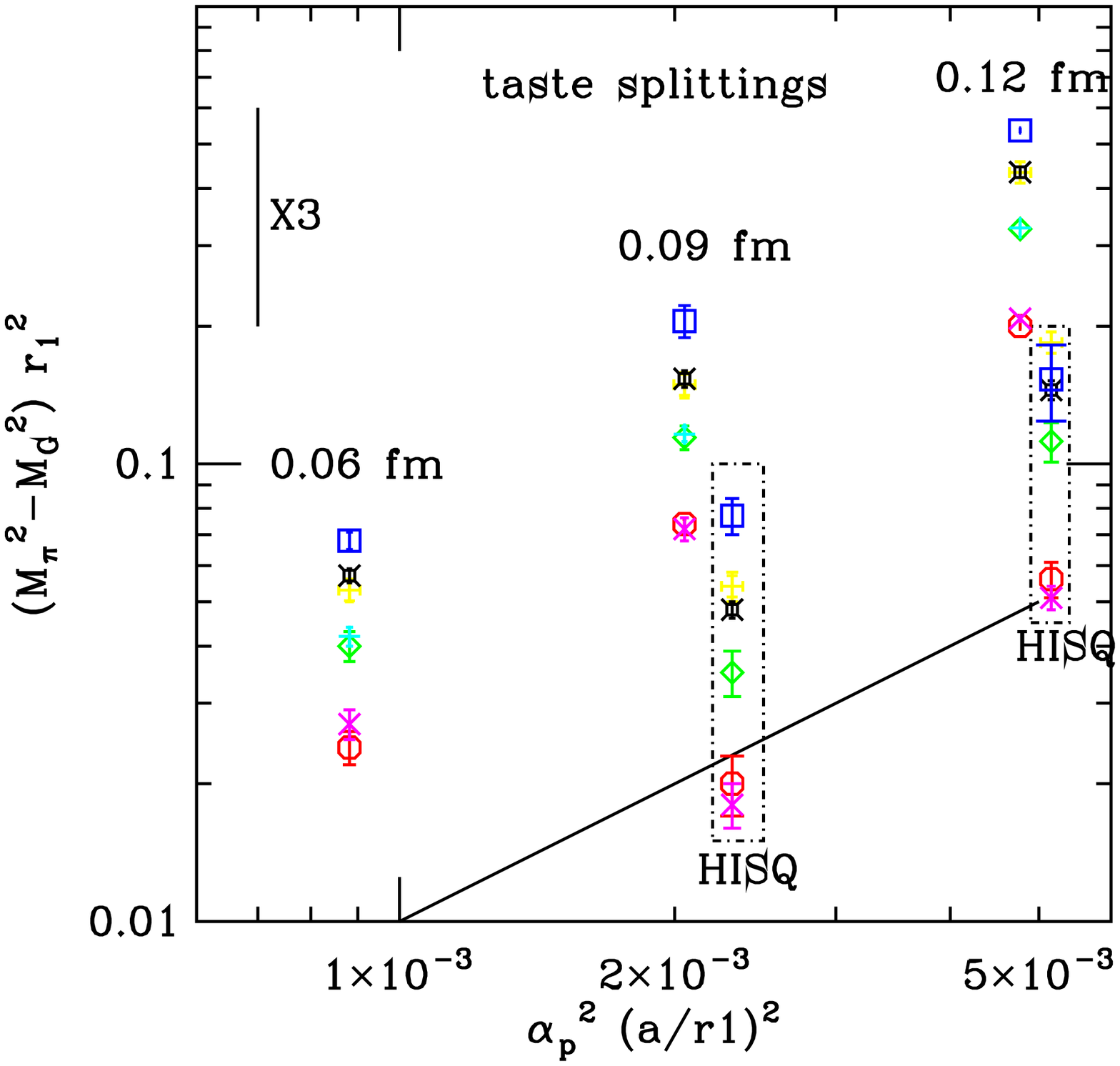}\hfill
    \end{minipage}
  \end{tabular}
\caption{Left: Sketch of phase diagram at zero baryon density as a
  function of light quark masses $m_u = m_d$ and $m_s$.  Right:
  Splitting of the pion taste multiplet showing the expected decrease
  with lattice spacing.  The unboxed points are for the asqtad action
  and the much lower boxed points are for the HISQ action.
\label{fig:phase_diag}
}
\end{figure}

Figure~\ref{fig:phase_diag} (left) summarizes in qualitative terms our
knowledge of the QCD phase diagram as a function of the light (up,
down, and strange) quark masses $m_u = m_d$ and $m_s$.  Lattice
calculations aim to check this picture.  One impotant question is
whether, when we fix the strange quark mass at its physical value and
reduce the up and down quark masses, we encounter the first-order
transition region.  Present indications are that we do not.

What can we learn about QCD thermodynamics from numerical lattice
simulations?  Here is a list of objectives.  More could be added.
\begin{itemize}
\item Obtaining accurate values of the crossover temperature $T_p$.
\item Determining the equation of state, velocity of sound, etc.
\item Studying critical universality at low light quark masses.
\item Calculating transport properties of the plasma.
\item Establishing the extent of validity of the hadron resonance gas
      model at low $T$.
\item Determinining the behavior of in-medium hadronic modes
      (e.g. $J/\psi$), especially above $T_p$?
\item Searching for an experimentally accessible critical
      point at nonzero baryon number density.
\end{itemize}
For all of these topics a nonperturbative treatment is necessary.
Numerical simulation on the lattice gives a first-principles,
nonperturbative treatment. We know of no alternative.  It does not
answer all of our questions, however.  Here is a list of limitations:
\begin{itemize}
  \item We can treat only static thermodynamic equilibrium or small
    perturbations around it.
  \item We work in euclidean time: Real time properties are difficult
    to extract. Transport properties can be computed, in principle,
    but it is not easy.
  \item Calculations at nonzero quark number density are very difficult.
\end{itemize}
Phomenological models can help extrapolate from lattice results to
regimes that are inaccessible to lattice calculations.

\section{Lattice Methodology}

For an introduction to lattice methods for QCD thermodynamics, please
see \cite{DeTar:2009ef} and references therein.  Here we mention only
a few key concepts.

\subsection{Feynman path integral}

We work with quantum grand-canonical partition function
\be
Z = \Tr \left[\exp\left(-H/T + \sum_i \mu_i N_i/T\right)\right],
\ee
for temperature $T$, QCD hamiltonian $H$, chemical potential $\mu_i$,
and conserved charge $N_i$.  It is rewritten, using the Feynman path
integral approach, as the functional integral
\be
Z = \int dA_\mu\, d\psi\, d\bar\psi \, \exp[-S(A,\psi,\bar\psi,\mu)]
\ee
where $A_\mu$, $\psi$, $\bar \psi$ represent the gluon and quark
fields and $S$ is the classical action in a Eudlidean space-time
(imaginary time).  The continuous space-time is discretized as a
lattice of points of spacing $a$, and the classical action is
formulated on that lattice.  The parameters of the action are, as
usual, the gauge coupling and the quark masses.  Introducing the
lattice puts the functional integration in a form that is more
amenable to numerical simulation, and it provides the ultraviolet
regulation needed to define QCD.

\subsection{Varying the temperature}

The imaginary time coordinate has a finite extent determined by the
temperature.  So if there are $N_\tau$ points in the time direction,
at lattice spacing $a$, the temperature is given by $T = 1/(a N_\tau)$.
There are two methods in current use for varying the temperature.
\begin{enumerate}
\item Fixed $N_\tau$ method.  Through the renormalization group, the
  lattice spacing $a$ depends on the bare gauge coupling $g$, so as
  $g$ decreases, $a$ decreases, and $T$ increases.  Low $T$ then
  implies larger lattice spacing and larger cutoff effects!  With this
  method we scan a temperature range at one fixed $N_\tau$ and then
  repeat at larger $N_\tau$ to move closer to the continuum.
\item Fixed scale method. \cite{Levkova:2002sv,Umeda:2008bd} 
  With this method we fix the gauge
  coupling and lattice spacing and vary $N_\tau$.  Cutoff effects are
  then uniform in $T$.
\end{enumerate}

\subsection{Setting the bare quark masses}

Quark masses can also be varied to explore the phase diagram.  It is
useful to work along ``lines of constant physics''; {\it i.e.}  we
tune the bare quark masses so as to keep (zero-temperature) meson
masses fixed in physical units as $T$ (so $a$) is varied.  Typically
we set the strange quark mass $m_s$ to its physical value, but it is
expensive to calculate with a physical up and down quark mass $m_u
\approx m_d = m_\ell$, so we fix the ratio $m_\ell/m_s$, repeat the
calculation for a range of ratios, and then extrapolate to the
physical point.

\subsection{Determining the lattice scale}

To get $T$ in MeV we need to know $a$ in physical units.  This value
is determined in a zero temperature calculation at the same
hamiltonian parameters.  It requires matching one dimensionful lattice
result with one experimental result.  Two common methods are in use:
  \begin{enumerate}
  \item $f_K$ scale.  One measures the meson decay constant in lattice
    units $a f_K$ at zero temperature.  From the experimental value of
    $f_K$, we then know $a$.
  \item $r_1$ or $r_0$ method.  This method is based on a measurement
    of the static quark-antiquark potential, a relatively easy
    process.  The constant $r_1$ is defined as the value of $R$ where
    $R^2 dV(R)/dR = 1$.  The Sommer scale $r_0$ is similarly defined
    \cite{Sommer:1993ce}.  Of course, these values are not measured in
    experiment.  So their values are determined in terms of an
    experimentally observable quantity, such as the splitting of the
    $\Upsilon$ spectrum, with the result $r_1 \approx 0.31$ fm and
    $r_0 \approx 0.47$ fm \cite{Bazavov:2009bb}.
  \end{enumerate}
All scale definitions must agree at zero lattice spacing and physical
quark masses, but we expect some disagreement at nonzero spacing and
unphysical masses.  With current methods we can get better than $\sim
2$\% accuracy in $T$.

\subsection{Lattice fermion doubling problem}

Putting fermions on the lattice is nontrivial.  Discretization of the
Dirac action introduces complications.  As a result there are several
lattice fermion formulations, each with its advantages and
disadvantages.  With a naive discretization in three space and one
time dimension we get $2^4$ quark species of the same mass.  This is
called the fermion ``doubling'' problem.  The remedial strategy varies
with the fermion implementation.  

Wilson introduced a dimension-five term in the action to lift the
degeneracy.  All unwanted fermions then get masses of order $1/a$.
This procedure breaks chiral symmetry explicitly, which adds to the
complexity of studies at finite temperature.

The domain wall and overlap implementations usually start from
Wilson's action and build from it an action with a form of chiral
symmetry.  It is rigorous, elegant, but computationally expensive.

The staggered fermion implementation diagonalizes the fermion matrix
partially to reduce the degeneray from 16 to 4.  In modern language,
these are called ``tastes''.  (Then each flavor comes in four tastes.)
Finally, one takes the fourth root of the fermion determinant to get
an approximately correct counting of sea quark flavors.  This is a
controversial step, but recent work has placed it on firmer
theoretical ground. (See a discussion and references in
\cite{Bazavov:2009bb}.)

The lattice regulates ultraviolet divergences by introducing a
momentum cutoff of order $1/a$.  As the spacing is reduced, we remove
the cutoff.  Depending on how the lattice action is formulated, at
nonzero lattice spacing, results can be distorted by the cutoff.  The
goal of improving the formulation is to reduce these effects at a
given $a$.  This is done by adding irrelevant higher-dimensional terms
to the action \cite{Symanzik:1983gh}. The original staggered fermion
action is ``unimproved'': good to ${\cal O}(a^2)$.  Improved
formulations in current wide use are called ``p4''
\cite{Heller:1999xz,Karsch:2000ps}, ``asqtad'' (for references, see
\cite{Bazavov:2009bb}), ``stout'' \cite{Morningstar:2003gk,Aoki:2005vt}, and
``HISQ'' \cite{Follana:2006rc}.

In the continuum limit the tastes are described by an exact,
extraneous SU(4) symmetry, and the fourth root is trivial.  At nonzero
lattice spacing, this symmetry is broken, which leads to a distortion
of the hadron spectrum, as shown below. 
As we will see, recent calculational results suggest that taste
symmetry breaking is the source of a large share of the cutoff effects
in traditional staggered fermion thermodynamics.  Currently, the HISQ
action has the most improved taste symmetry, followed closely behind
by stout, and then asqtad, and p4.  Aside from taste splitting, other
cutoff effects are expected.  The p4, asqtad, and HISQ actions are all
improved with leading errors at ${\cal O}(a^2 \alpha_s)$, and the
stout action is less improved with leading errors at ${\cal O}(a^2)$.

The effects of taste-symmetry breaking are most evident in the pion
spectrum.  Four tastes of quarks and four of antiquarks yield a
multiplet of sixteen pion tastes for each physical pion.  The
resulting multiplet structure is shown in Fig.~\ref{fig:phase_diag}
\cite{Bazavov:2010ru}.  The figure shows that the splitting decreases
approximately as $a^2 \alpha_V^2$.  The considerable improvement of
HISQ over asqtad is also apparent.

\section{Results}

I will review some recent results mostly from the HotQCD
collaboration \footnote{ A.~Bazavov, T.~Bhattacharya, M.~Cheng,
  N.H.~Christ, C.~DeTar, S.~Gottlieb, R.~Gupta, U.M.~Heller, C.~Jung,
  F.~Karsch, E.~Laermann, L.~Levkova, C.~Miao, R.D.~Mawhinney,
  S.~Mukherjee, P.~Petreczky, D.~Renfrew, C.~Schmidt, R.A.~Soltz,
  W.~Soeldner, R.~Sugar, D.~Toussaint, W.~Unger and P.~Vranas. }
including some very new ones based on the HISQ action $N_\tau = 6,8$
and asqtad $N_\tau = 12$.

\begin{figure}[t]
\centering
\includegraphics[width=0.32\textwidth]{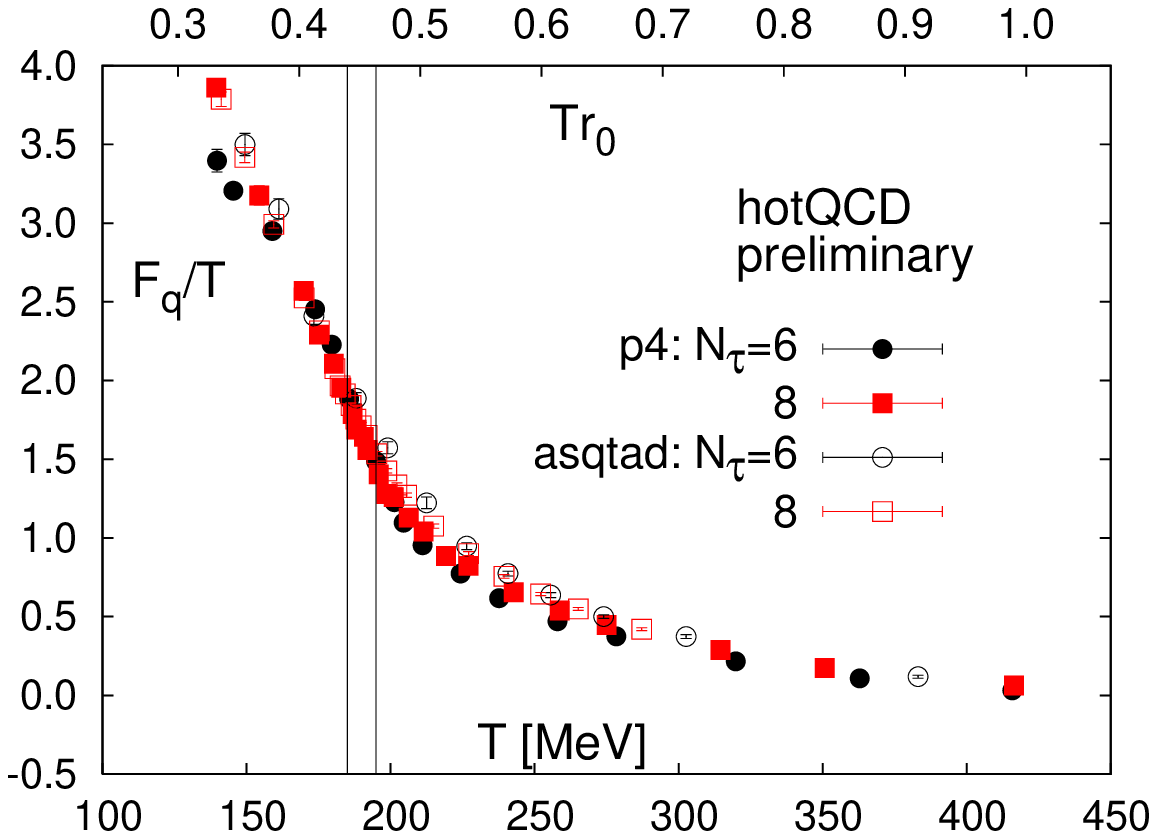}
\includegraphics[width=0.30\textwidth]{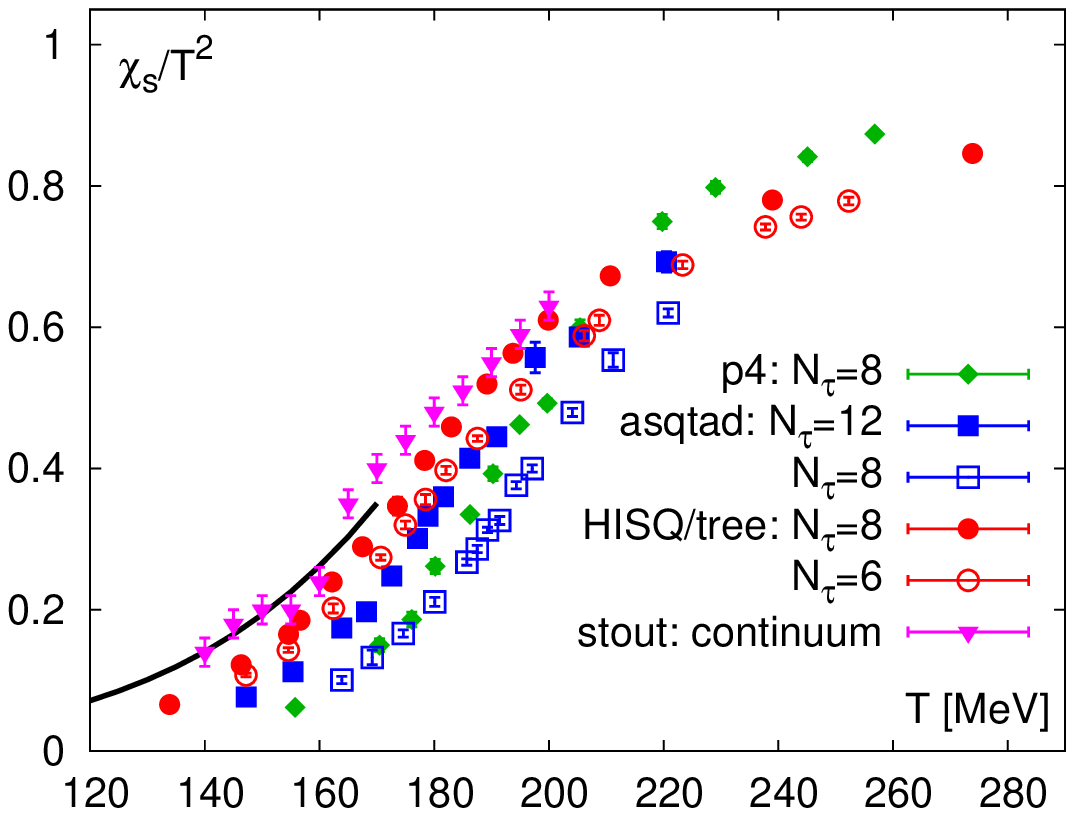}
\includegraphics[width=0.33\textwidth]{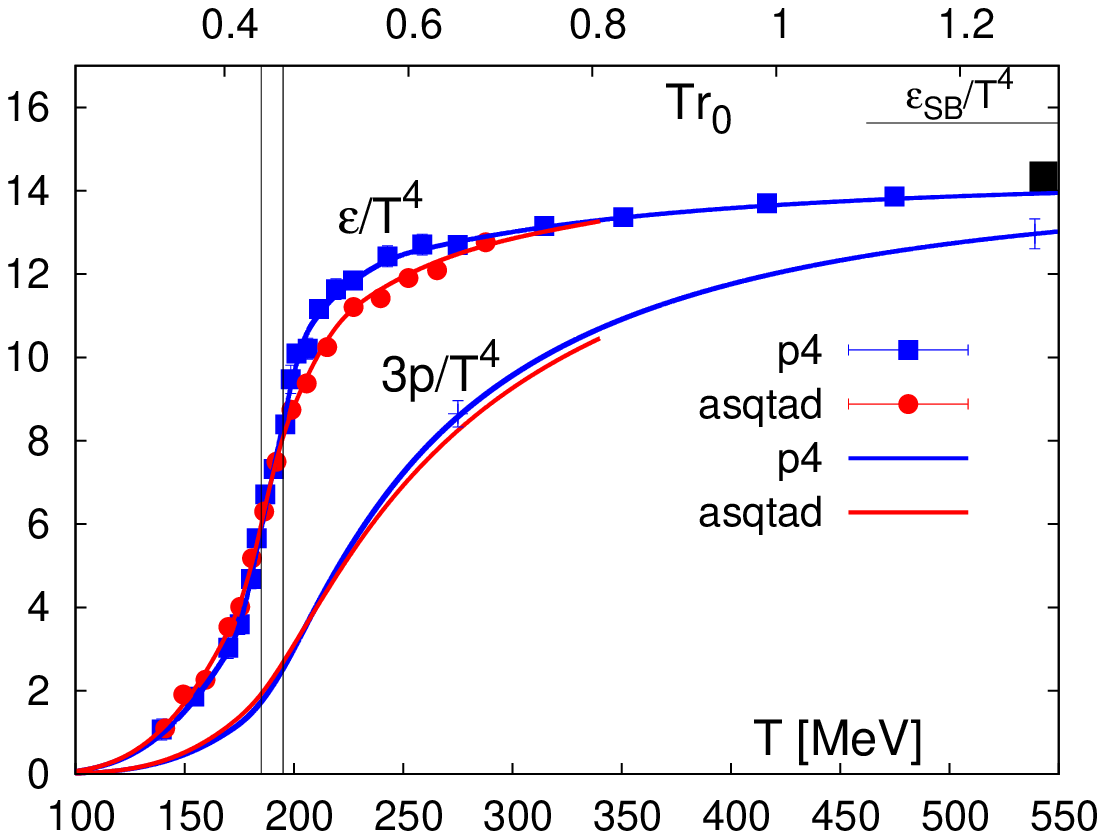}
\caption{Three deconfinement markers as a function of temperature (MeV).
  Left: static quark free energy \protect\cite{DeTar:2009ef}. 
  Middle: strange quark number susceptibility \protect\cite{Bazavov:2010ru}.
  Right: energy density and pressure in units of temperature 
  \cite{Bazavov:2009zn}.
  The Stefan-Boltzmann free-gas limit is indicated on the right. 
  The crossover is evident in all of them.
\label{fig:deconf}
}
\end{figure}

\subsection{Indicators of deconfinement}

A variety of observables are good phenomenological indicators of
deconfinement.  We discuss two of them, namely the Polyakov loop or
``static quark self energy'' and the strange quark number
susceptibility.  A third, the equation of state, is discussed later
below.

The traditional deconfinement indicator is the ``Polyakov loop'' $L$.
It is related to the static quark free energy $F_q$, {\it i.e.} the
difference of the free energy of the thermal ensemble with and without
a static quark:
\be
  L = \VEV{\Tr P \exp(ig\int_0^{1/T}d\tau\, A_0(\tau))} \sim \exp[-F_q(T)/T]
\ee
Even when light quarks are present in the ensemble, adding a static
quark at low temperature requires screening by a light quark,
increasing the free energy by an amount equal, roughly, to a
constituent quark mass.  In the deconfined phase the constitutent
quark mass is very low.  This effect is visible in the left panel of
Fig.~\ref{fig:deconf}.
There is no direct linkage between this quantity and the chiral order
parameter, so this observable is not a good indicator of the chiral
transition.

The strange quark number susceptibiltiy measures fluctuations in
strangeness $\chi_s = \VEV{S^2}/(VT)$.  Such fluctuations are expected
to be large in the deconfined phase where strangeness is carried by
the quark degrees of freedom, and small in the confined phase where
it is carried by hadrons containing a strange quark.  This behavior is
apparent in the right panel of Fig.~\ref{fig:deconf}.  Although this
quantity is expected to have a singularity at the chiral critical
point, an analysis of critical behavior suggests that the singularity
is too mild to make this observable a good indicator of the chiral
transition.

\subsection{Indicators of chiral symmetry restoration}

\begin{figure}
  \begin{tabular}{lcr}
    \begin{minipage}{0.306\textwidth}
    \includegraphics[width=\textwidth]{figs/pbp_nf21_nt8.ps}\hfill
    \end{minipage}
&
    \begin{minipage}{0.297\textwidth}
    \vspace*{-3mm}
    \includegraphics[width=\textwidth]{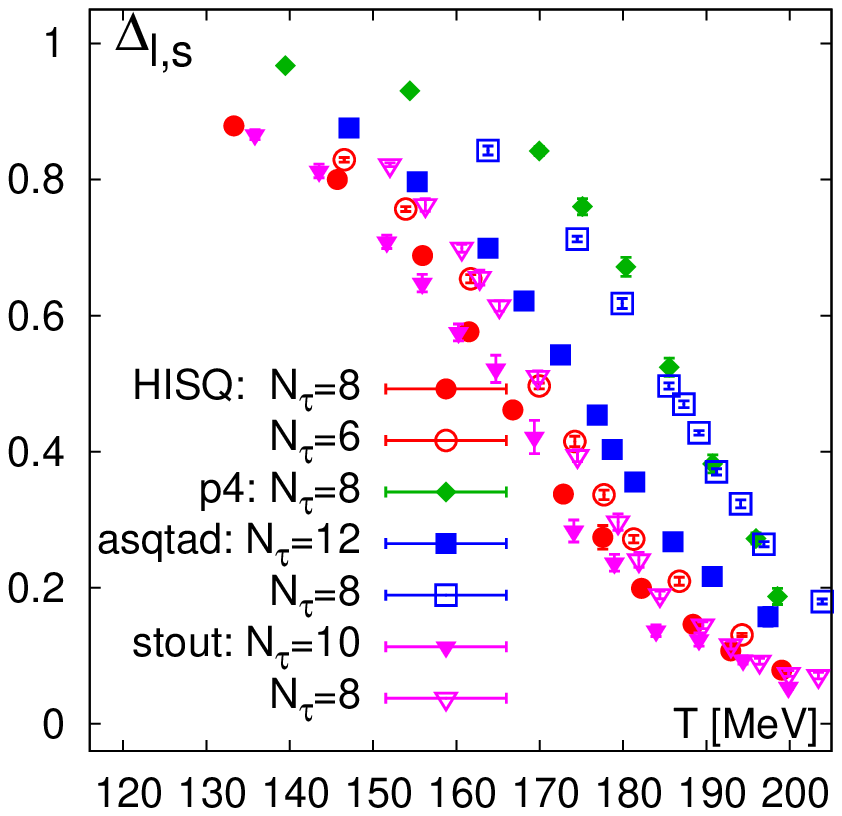}\hfill
    \end{minipage}
&
    \begin{minipage}{0.35\textwidth}
    \vspace*{-3mm}
    \includegraphics[width=\textwidth]{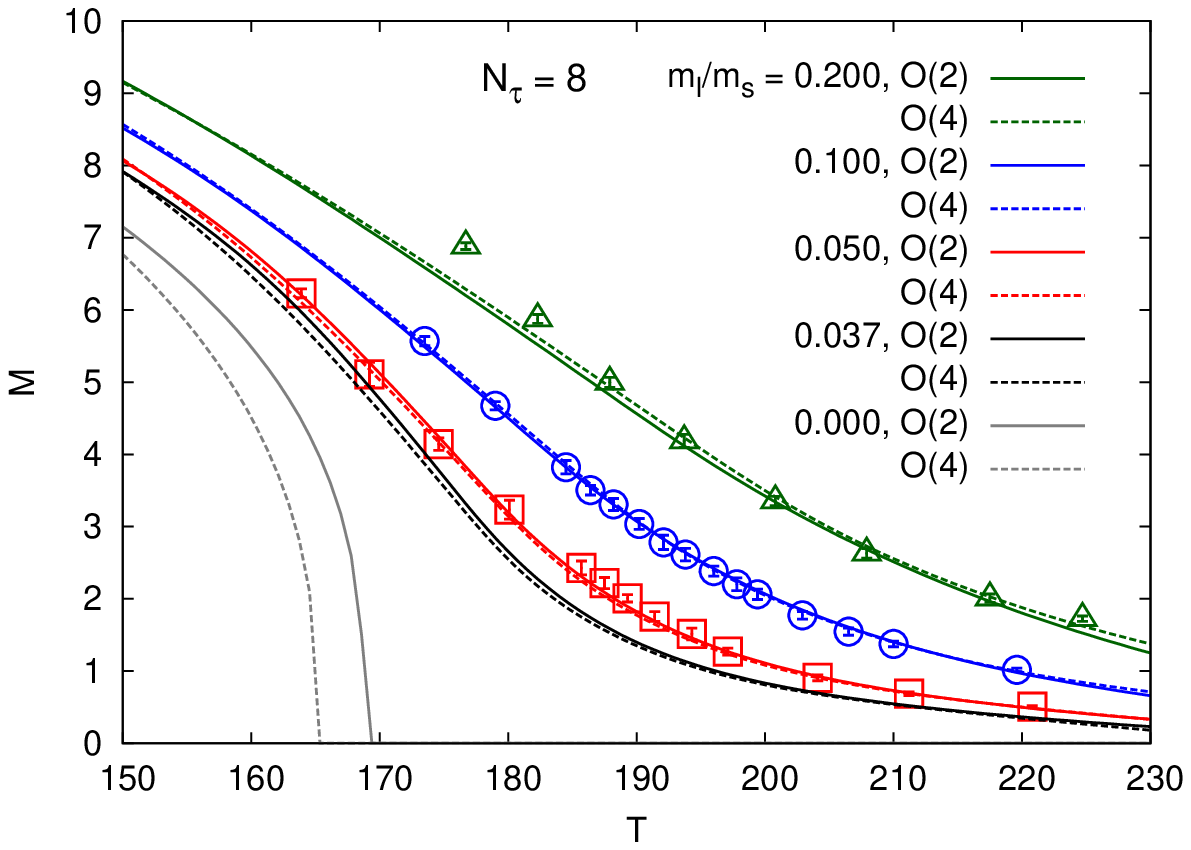}
    \end{minipage}
  \end{tabular}
\centering
\caption{Left: Chiral condensate for a variety of quark mass ratios
  $m_\ell/m_s$ \protect\cite{Bernard:2004je}.  Also shown is the
  extrapolation to zero quark mass where a singularity is expected to
  appear.  Middle: The subtracted chiral condensate (see text) from
  various fermion formulations, showing a lowering of the transition
  temperature with decreased taste splitting \protect\cite{Bazavov:Lat2010}.
  Right: Chiral condensate for the asqtad action fit to the O(2) and
  O(4) critical scaling functions (see text)
  \protect\cite{Bazavov:Lat2010}.
\label{fig:pbp_comp}}.
\end{figure}

The chiral condensate and its associated susceptibility are obvious
markers of chiral symmetry restoration.  The light quark chiral
condensate $\pbp$ is, in fact, the order parameter for chiral symmetry
at zero up and down quark masses.
\be
  \VEV{\pbp} = (T/V)\partial \log Z/\partial m.
\ee
It is nonzero when chiral symmetry is spontaneously broken and zero
when it is restored.  We expect restoration at high $T$.  When all sea
quark masses are nonzero, chiral symmetry is not exact, so we don't
get zero, exactly.  The example in Fig.~\ref{fig:pbp_comp} (left)
confirms the expected behavior.

The chiral condensate is subject to both additive (divergent at
nonzero quark mass) and multiplicative renormalization.  To compare
results for different actions, it is necessary to remove these
factors.  A convenient choice is the ``subtracted'' condensate (middle
panel of Fig.~\ref{fig:pbp_comp}):
\be
  \Delta_{\ell,s} = 
  [\VEV{\pbp}_\ell(T) - m_\ell/m_s \VEV{\pbp}_s(T)]/
  [\VEV{\pbp}_\ell(T=0) - m_\ell/m_s \VEV{\pbp}_s(T=0)]
\ee

\subsection{Taste symmetry and the transition temperature}

In Fig.~\ref{fig:deconf} and Fig.~\ref{fig:pbp_comp} we see that the
various actions give strikingly different.  The discrepancies
correlate with the degree of taste symmetry of the action.  As taste
symmetry is improved, the curves shift to lower temperature.  This
is achieved by decreasing the lattice spacing, {\it i.e.},
increasing $N_\tau$ and by improving the action.  For the latter
property, in order of gradually improved taste symmetry, the actions
are p4, asqtad, stout, and HISQ.

\subsection{Scaling of chiral order parameter (Magnetic equation of state)}

At zero quark mass we expect universal $O(4)$ critical behavior at the
chiral-symmetry-restoring phase transition.  It is $O(2)$ at nonzero
lattice spacing for staggered fermion actions. Define 
\be
t = (T -T_c)/T_c \ \ \ \mbox{and} \ \ \ h = (m_\pi/m_K)^2 \approx m_\ell/m_s
\ee
For small $h$ and $t$ we have
\be
  M(t,h) \equiv m_s/T^4 \VEV{\pbp(t,h)} \rightarrow t^{\beta} f(z) + {\rm regular}
\ee
where $z = z_0 t/h^{1/(\beta\delta)}$ and $f(z)$ is the universal
scaling function for $O(2)$ or $O(4)$.  This analysis is tested in
Fig.~\ref{fig:pbp_comp} (right).  It follows the analysis for the p4
action described in \cite{Ejiri:2009ac}.  A similar analysis for
Wilson fermions long ago found surprisingly good scaling
\cite{Iwasaki:1996ya}.  Such a scaling analysis gives a framework for
extrapolating results to the physical quark mass.

\subsection{Chiral susceptibility}

\begin{figure}
\centering
\includegraphics[width=0.41\textwidth]{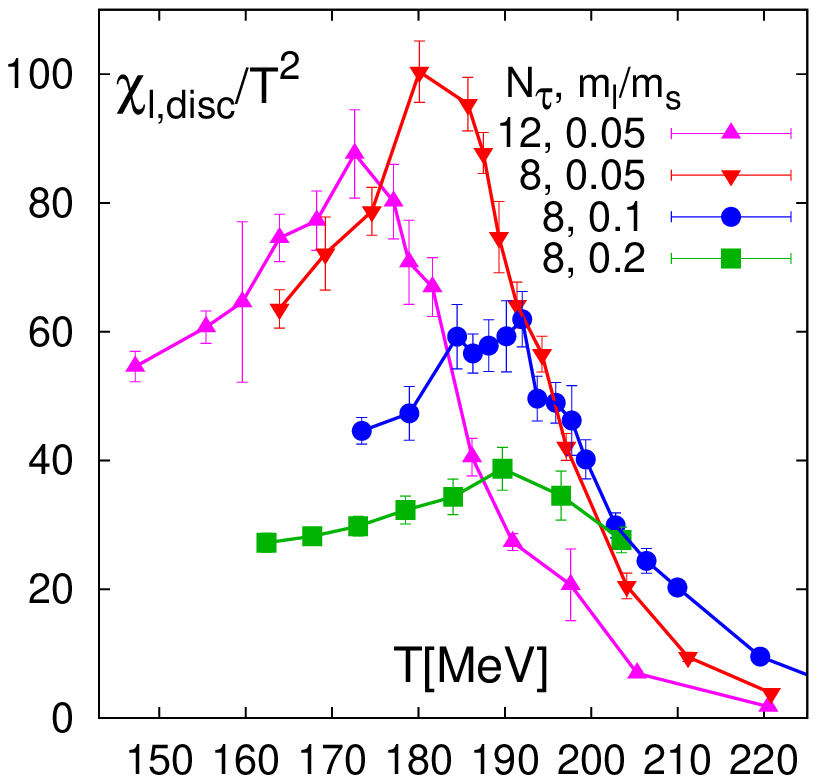}
\includegraphics[clip=on,width=0.39\textwidth]{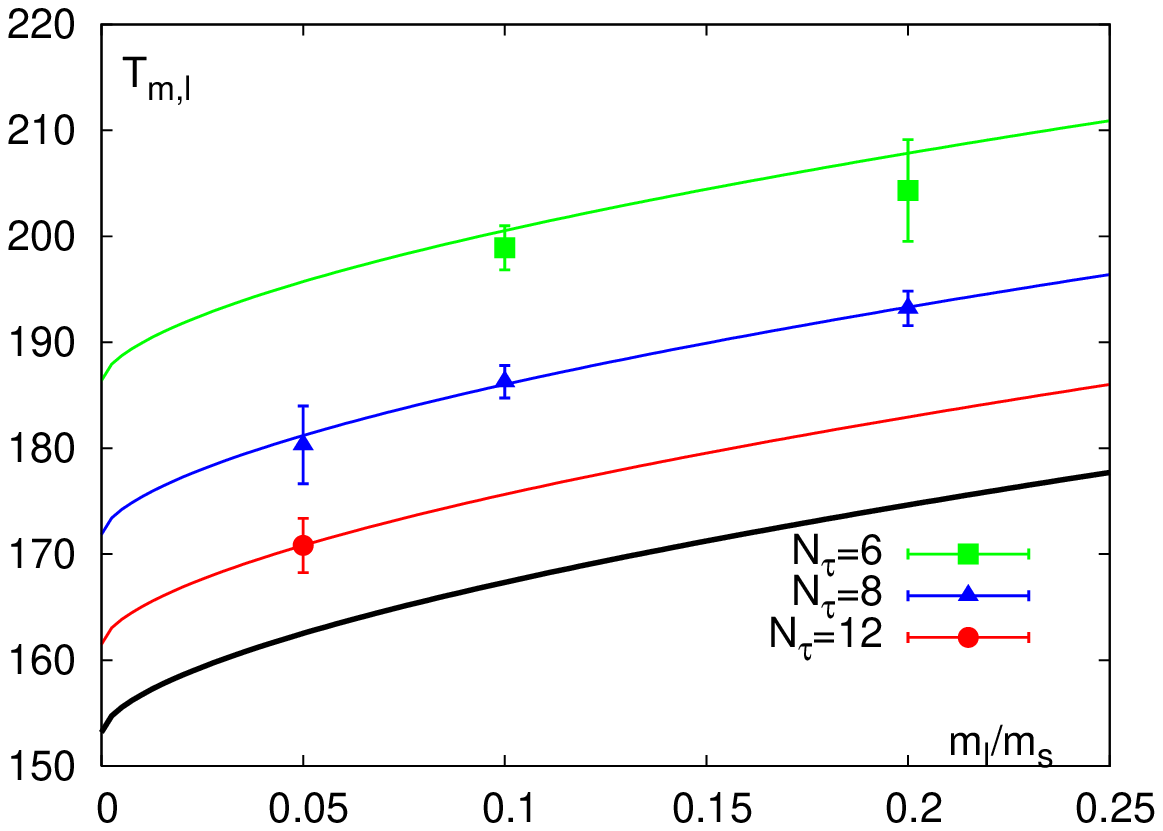}
\caption{Left: Disconnected light quark chiral susceptibility for the
  asqtad action showing a peak at the crossover temperature 
  \protect\cite{Soeldner:Lat2010}.  The peak
  shifts to lower temperatures with increasing $N_\tau$ (decreasing
  lattice spacing).  Right: The crossover temperature as a function of
  the light quark mass ratio and $N_\tau$ \protect\cite{Soeldner:Lat2010}.  
  The curves show the fit to
  Eq.~(\protect\ref{eq:tc_lt}).
\label{fig:susc}
}
\end{figure}

The chiral susceptibility measures fluctuations in the chiral
condensate.  For light quarks it is

\be
   \chi_\ell = \frac{T}{V}\frac{\partial^2}{\partial m_\ell^2}\log Z
          = \chi_{\ell,disc} + 2 \chi_{\ell,conn}.
\ee
The ``disconnected'' and ``connected'' labels refer to the topology of
quark world lines in the conventional computation.  The disconnected
term peaks at the crossover, as shown in Fig.~\ref{fig:susc}.  The
peak height diverges in the chiral limit.  Thus it is an excellent
marker for the crossover.  Consistent with the behavior of chiral
condensate, the peak shifts to lower temperature as the lattice
spacing is decreased (increasing $N_\tau$).

\subsection{Transition temperature at the physical point }

Locating the peaks of the chiral susceptibility at other quark masses
and lattice spacings allows us to carry out an extrapolation to the
physical light quark mass (approximately $m_s/27$) and zero lattice
spacing.  The temperature at the peak is plotted in
Fig.~\ref{fig:susc} (right) together with curves based on the ansatz
\be
T_p = T_c(0) + a (m_\ell/m_s)^{1/(\beta\delta)} + b/N_\tau^2
\label{eq:tc_lt}
\ee
The light quark mass dependence is motivated by the expected universal
O(4) critical behavior ($1/(\beta\delta) = 0.54$) and the lattice
spacing ($1/N_\tau$) dependence is based on the expected $O(a^2)$
cutoff dependence of the action.

At the physical point $m_\ell/m_s = 1/27$ and zero lattice spacing we
obtain a preliminary value of the crossover temperature at the
physical point: $T_p({\rm phys}) \approx 164(6)$ MeV
\cite{Soeldner:Lat2010}. The Budapest-Wuppertal result for a closely related
observable is 147(2)(3) MeV \cite{Borsanyi:2010bp}

In the past couple of years there has been a lively discussion about
the transition temperature.  In 2004 the MILC collaboration, using the
improved asqtad action, carried out a similar extrapolation from
$N_\tau = 4$, 6, and 8 to the physical point with lower statistics
than in the present study and reported 169(12)(4) MeV
\cite{Bernard:2004je}.  In 2006 Cheng {\it et al.}, using the p4
action, reported 192(7)(4) MeV at the physical point based on
simulations at $N_\tau = 4$ and 6 \cite{Cheng:2006qk}.  The HotQCD
collaboration published a study of the equation of state in 2009,
based on both the asqtad and p4 actions, but, because there were not
enough data to do so at the time, quite deliberately did not quote a
result for the transition temperature at the physical point
\cite{Bazavov:2009zn}.  At the same time the Budapest-Wuppertal
collaboration reported on its study using the stout action, with
several values depending on the observable, including 147(2)(3) MeV
from their renormalized disconnected susceptibility and 165(5)(3) from
the strange quark number susceptibility \cite{Borsanyi:2010bp}.

What we have learned first from these studies is that the transition
temperature is more sensitive to taste-breaking effects in the
staggered action than some had expected.  But the story is not
finished.  The HotQCD collaboration has undertaken a more
comprehensive analysis of O(N) universality with its current data.
This study may lead to a more refined determination of the crossover
temperature. It also provides a means of deciding which observables
are better markers of critical behavior.

\subsection{Equation of state (trace anomaly)}

\begin{figure}
\centering
\includegraphics[width=0.45\textwidth]{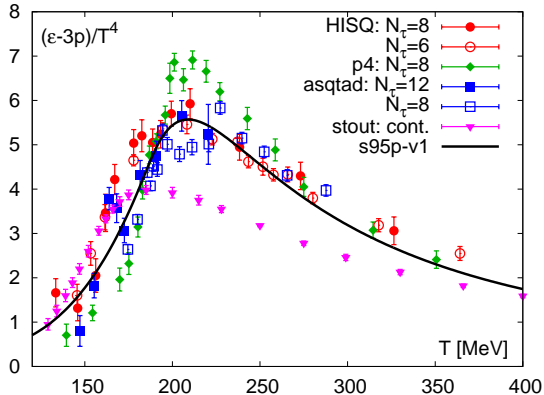}
\hfill
\caption{Interaction measure in units of temperature {\it vs.}
  temperature for the asqtad, p4, HISQ, and stout actions at a variety
  of values of $N_\tau$ \protect\cite{Bazavov:Lat2010}.  The curve
  is a convenient parameterization.  It agrees with the hadron
  resonance gas model at low temperature.
\label{fig:e-3p}
}
\end{figure}

The equation of state, {\it i.e.}, the energy density $\epsilon$,
pressure $p$, and entropy density $s$ as a function of temperature is
an important quantity in the hydrodynamics of heavy ion collisions and
in the characterization of the early universe.  The now standard
lattice QCD construction of the equation of state begins with a
calculation of the ``trace anomaly'' or ``interaction measure'', $I =
\epsilon - 3p$.  It is plotted in Fig.~\ref{fig:e-3p} for a variety of
actions.
At low temperature the measured points lie below the prediction of the
hadron resonance gas model (based on physical hadron masses).  Such an
effect is an expected consequence of the splitting of the pion taste
multiplet, which tends to increase the rms mass of the pion, and
therefore increase the transition temperature.  At high temperature
where cutoff effects are much reduced, the three actions (p4, asqtad,
and HISQ) agree.  However, a recent Budapest-Wuppertal result for the
stout action shows a significant deviation \cite{Borsanyi:2010cj}.
The results are compared in Fig.~\ref{fig:e-3p}.  The stout action
points include a rather large ``tree-level'' correction for cutoff
effects.  Since the p4, asqtad, and HISQ actions are improved at
${\cal O}(a^2)$, they have better scaling properties at high
temperature, as shown, and no such correction was applied.  In any
case the correction vanishes for all actions in the continuum limit.

The pressure $p$ and energy density $\epsilon$ are obtained from the
interaction measure $I$ as follows:
\be
    p = \frac{T}{V}\int_{T_0}^T dT^\prime\, \frac{1}{T^{\prime 5}}I(T^\prime) \ \ \ \  \epsilon = I + 3p 
\ee
Results are shown on the right in Fig.~\ref{fig:deconf} for the asqtad
and p4 actions.

\section{Conclusions}

  Lattice QCD is providing a wealth of information about high
  temperature QCD, particularly about the nature of the transition
  from low to high temperature and the behavior of several quantities
  of phenomenological importance, including the equation of state and
  the quark number susceptibility.  Other quantities I did not have
  space to discuss are the speed of sound, the equation of state at
  nonzero baryon density, transport properties, and the survival of
  hadronic modes in the medium.

  The staggered fermion formulation is most widely used for
  thermodynamic studies.  We have learned that taste-symmetry breaking
  makes a large contribution to cutoff effects in that formulation
  making it highly desirable to use actions such as HISQ and stout
  that have better taste symmetry.

  More is yet to be learned about the critical scaling of various
  quantities near the chiral phase transition, and further study is
  needed to settle substantial disagreements in the interaction
  measure at moderate temperature.

% Non-BibTeX users please use


\begin{thebibliography}{99}
%
\bibitem{Hwa:2010}
  R.~Hwa and X.N.~Wang, ed., {\it Quark Gluon Plasma 4}, (World Scientific,
  Singapore, 2010).
\bibitem{Yagi:2008}
  K. Yagi, T. Hatsuda, and Y. Miake, ``Quark Gluon Plasma: 
  From big bang to little bang'' (Cambridge University Press, Cambridge,
  2008).
\bibitem{DeTar:2009ef}
  C.~DeTar and U.~M.~Heller,
  %``QCD Thermodynamics from the Lattice,''
  Eur.\ Phys.\ J.\  A {\bf 41}, 405 (2009)
  [arXiv:0905.2949 [hep-lat]].
  %%CITATION = EPHJA,A41,405;%%
\bibitem{Levkova:2002sv}
  L.~Levkova,
  %``Staggered fermion thermodynamics using anisotropic lattices,''
  Nucl.\ Phys.\ Proc.\ Suppl.\  {\bf 119}, 520 (2003)
  [arXiv:hep-lat/0209069].
  %%CITATION = NUPHZ,119,520;%%
\bibitem{Umeda:2008bd}
  T.~Umeda, S.~Ejiri, S.~Aoki, T.~Hatsuda, K.~Kanaya, Y.~Maezawa and H.~Ohno,
  %``Fixed Scale Approach to Equation of State in Lattice QCD,''
  Phys.\ Rev.\  D {\bf 79}, 051501 (2009)
  [arXiv:0809.2842 [hep-lat]].
  %%CITATION = PHRVA,D79,051501;%%
\bibitem{Sommer:1993ce}
  R.~Sommer,
  %``A New way to set the energy scale in lattice gauge theories and its
  %applications to the static force and alpha-s in SU(2) Yang-Mills theory,''
  Nucl.\ Phys.\  B {\bf 411}, 839 (1994)
  [arXiv:hep-lat/9310022].
  %%CITATION = NUPHA,B411,839;%%
\bibitem{Bazavov:2009bb}
  A.~Bazavov {\it et al.},
  %``Full nonperturbative QCD simulations with 2+1 flavors of improved staggered
  %quarks,''
  Rev.\ Mod.\ Phys.\  {\bf 82}, 1349 (2010)
  [arXiv:0903.3598 [hep-lat]].
  %%CITATION = RMPHA,82,1349;%%
\bibitem{Symanzik:1983gh}
  K.~Symanzik,
  %``Continuum Limit And Improved Action In Lattice Theories. 2. O(N) Nonlinear
  %Sigma Model In Perturbation Theory,''
  Nucl.\ Phys.\  B {\bf 226}, 205 (1983).
  %%CITATION = NUPHA,B226,205;%%
\bibitem{Heller:1999xz}
  U.~M.~Heller, F.~Karsch and B.~Sturm,
  %``Improved staggered fermion actions for QCD thermodynamics,''
  Phys.\ Rev.\  D {\bf 60}, 114502 (1999)
  [arXiv:hep-lat/9901010].
  %%CITATION = PHRVA,D60,114502;%%
\bibitem{Karsch:2000ps}
  F.~Karsch, E.~Laermann and A.~Peikert,
  %``The pressure in 2, 2+1 and 3 flavour QCD,''
  Phys.\ Lett.\  B {\bf 478}, 447 (2000)
  [arXiv:hep-lat/0002003].
  %%CITATION = PHLTA,B478,447;%%
\bibitem{Morningstar:2003gk}
  C.~Morningstar and M.~J.~Peardon,
  %``Analytic smearing of SU(3) link variables in lattice QCD,''
  Phys.\ Rev.\  D {\bf 69} (2004) 054501
  [arXiv:hep-lat/0311018].
  %%CITATION = PHRVA,D69,054501;%%
\bibitem{Aoki:2005vt}
  Y.~Aoki, Z.~Fodor, S.~D.~Katz and K.~K.~Szabo,
  %``The equation of state in lattice QCD: With physical quark masses  towards
  %the continuum limit,''
  JHEP {\bf 0601}, 089 (2006)
  [arXiv:hep-lat/0510084].
  %%CITATION = JHEPA,0601,089;%%
\bibitem{Follana:2006rc}
  E.~Follana {\it et al.}  [HPQCD Collaboration and UKQCD Collaboration],
  %``Highly Improved Staggered Quarks on the Lattice, with Applications to Charm
  %Physics,''
  Phys.\ Rev.\  D {\bf 75}, 054502 (2007)
  [arXiv:hep-lat/0610092].
  %%CITATION = PHRVA,D75,054502;%%
\bibitem{Bazavov:2010ru}
  A.~Bazavov {\it et al.}  [MILC collaboration],
  %``Scaling studies of QCD with the dynamical HISQ action,''
  Phys.\ Rev.\  D {\bf 82}, 074501 (2010)
  [arXiv:1004.0342 [hep-lat]].
  %%CITATION = PHRVA,D82,074501;%%
\bibitem{Iwasaki:1996ya}
  Y.~Iwasaki, K.~Kanaya, S.~Kaya and T.~Yoshie,
  %``Scaling of chiral order parameter in two-flavor QCD,''
  Phys.\ Rev.\ Lett.\  {\bf 78}, 179 (1997)
  [arXiv:hep-lat/9609022].
  %%CITATION = PRLTA,78,179;%%
\bibitem{Ejiri:2009ac}
  S.~Ejiri {\it et al.},
  %``On the magnetic equation of state in (2+1)-flavor QCD,''
  Phys.\ Rev.\  D {\bf 80}, 094505 (2009)
  [arXiv:0909.5122 [hep-lat]].
  %%CITATION = PHRVA,D80,094505;%%
\bibitem{Bazavov:Lat2010}
  A.~Bazavov and P.~Petreczky [HotQCD collaboration], (PoS LAT2010, to be published, 2010).
\bibitem{Soeldner:Lat2010}
  W.~S\"oldner [HotQCD collaboration], (PoS LAT2010, to be published, 2010).
\bibitem{Borsanyi:2010bp}
  S.~Borsanyi, Z.~Fodor, C.~Hoelbling, S.~D.~Katz, S.~Krieg, C.~Ratti 
  and K.~K.~Szabo [Wuppertal-Budapest Collaboration],
  %``Is there still any Tc mystery in lattice QCD? Results with physical masses
  %in the continuum limit III,''
  JHEP {\bf 1009}, 073 (2010)
  [arXiv:1005.3508 [hep-lat]].
  %%CITATION = JHEPA,1009,073;%%
\bibitem{Bernard:2004je}
  C.~Bernard {\it et al.}  [MILC Collaboration],
  %``QCD thermodynamics with three flavors of improved staggered quarks,''
  Phys.\ Rev.\  D {\bf 71}, 034504 (2005)
  [arXiv:hep-lat/0405029].
  %%CITATION = PHRVA,D71,034504;%%
\bibitem{Cheng:2006qk}
  M.~Cheng {\it et al.},
  %``The transition temperature in QCD,''
  Phys.\ Rev.\  D {\bf 74}, 054507 (2006)
  [arXiv:hep-lat/0608013].
  %%CITATION = PHRVA,D74,054507;%%
\bibitem{Bazavov:2009zn}
  A.~Bazavov {\it et al.},
  %``Equation of state and QCD transition at finite temperature,''
  Phys.\ Rev.\  D {\bf 80}, 014504 (2009)
  [arXiv:0903.4379 [hep-lat]].
  %%CITATION = PHRVA,D80,014504;%%
\bibitem{Borsanyi:2010cj}
  S.~Borsanyi {\it et al.},
  %``The QCD equation of state with dynamical quarks,''
  arXiv:1007.2580 [hep-lat].
  %%CITATION = ARXIV:1007.2580;%%
\end{thebibliography}
\end{document}